\documentclass[pre]{revtex4}
\usepackage{graphicx}
\usepackage{epsfig}
\begin{document}

\title{Dynamics of localized states in extended supersymmetric quantum mechanics with multi-well potentials}
\author{V.P. Berezovoj}
\email{berezovoj@kipt.kharkov.ua}
\author{M.I. Konchatnij}
\affiliation{A.I. Akhiezer Institute of Theoretical Physics, NSC ''KIPT'', 61108 Kharkov, Ukraine}

\begin{abstract}
In this paper we propose a self--consistent approach to the description of temporal dynamics of localized states. This approach is based on exactly solvable quantum mechanical models with multi-well potentials and their propagators. States of Hamiltonians with multi-well potentials form a suitable basis for the expansion of wave packets with different shapes and localization degrees. We also consider  properties of the tunneling wave packets, taking into account all states of Hamiltonians with symmetric and asymmetric potentials, as well as their dependence on the degree of localization and deformations of potentials. The study of the dynamics of initially localized states shows that application of the two-state approximation for the description of tunneling is considerably limited, especially for systems, which have several states in the under-barrier region, as for example in modern superconducting quantum interference devices and traps for cold atoms.
\end{abstract}
\maketitle

\section{Introduction}

Being one of the most exciting manifestations of the wave properties of particles, tunneling seemed to be a paradoxical prediction of quantum mechanics. Historically, the theory of the nuclear $\alpha$--decay was the first study of tunnel transitions \cite{susy-article-ref:1}, caused by bonding between nuclear resonances and continuous spectrum. Explanation of the intermolecular rearrangement of the ammonium spectrum \cite{susy-article-ref:2} due to the tunnel splitting of vibrational spectra, opened the era of study of tunneling in multi-well potentials \cite{susy-article-ref:3}. This problem became even more important in view of interpretation of the observed tunneling phenomena in condensed matter \cite{susy-article-ref:4,susy-article-ref:5}, and in particular in the Josephson junctions \cite{susy-article-ref:6}. Currently, investigations of tunneling processes are mostly related to the study of the Bose-Einstein condensate in different types of traps \cite{susy-article-ref:7,susy-article-ref:8}.

Theoretical analysis of tunneling in  double-well potentials is mostly performed within the two--mode approximation \cite{susy-article-ref:3}. Important characteristics of this approach are the difference of energies of the ground state and of the first excited state ($\Delta =E_{1} -E_{0} $), as well as their wave functions. Values of $\Delta $ define, in particular, revival times of the wave packets. This approximation explains general properties of tunneling, but it is unable to give interpretations of many subtle effects. Analysis of processes in multi-well potentials is complicated, since most of the  models use phenomenological or piecewise potentials (e.g., building from the rectangular wells and barriers, or from parabola), which are far from the real potentials. Spectra and wave functions in such potentials are unknown,  that supposes the numerical analysis of their properties in consequent studies. It is important to note that there exist exactly solvable models with multi-well potentials \cite{susy-article-ref:9, susy-article-ref:10, susy-article-ref:11, susy-article-ref:12}, which could be used for the description of tunneling processes.

Going beyond the two-mode approximation reveals interesting features of the dynamics of localized states. In particular, it leads to the substantial coherence breaking, as in symmetric \cite{susy-article-ref:13} as well as in asymmetric double-well potentials \cite{susy-article-ref:14, susy-article-ref:15, susy-article-ref:16}. Moreover, characteristics of tunneling strongly depend on the shape of the potential, and demonstrate a non-regular behavior which realized in bouncing increasing the probability of the localization of the wave packet (WP) in certain well. These features become even be more striking upon the initial localized state squeezing.

In QM the time evolution of wave packets is described by propagators \cite{susy-article-ref:20}, in which the contribution of the entire spectrum of the considered Hamiltonian is taken into account. However, only few models, mostly with one-well potentials, are known to have analytical expressions for propagators. To build propagators for exactly solvable quantum mechanical models with double-well potentials one could use the same approach as in \cite{susy-article-ref:10, susy-article-ref:17, susy-article-ref:18} and construct new propagators from the known ones.

The aim of the present paper is to describe the dynamics of localized states in multi-well potentials within the self--consistent approach. We use multi-well potentials obtained in the framework of $N=4~SUSY~QM$ \cite{susy-article-ref:12} and describe the dynamics of wave packets with corresponding propagators, calculated by use of the approach of \cite{susy-article-ref:10, susy-article-ref:17, susy-article-ref:18}.

The structure of this paper is as follows. In Section 1 we briefly discuss the construction of exactly solvable models in the framework of $N=4~SUSY~QM$ \cite{susy-article-ref:12} and give expressions for Hamiltonians with both symmetric and asymmetric potentials, and wave functions, obtained from the initial model of harmonic oscillator (HO). In Section 2 we obtain the expressions for propagators in these models using the approach of \cite{susy-article-ref:10, susy-article-ref:17, susy-article-ref:18}. Using the Hamiltonian of HO as the initial one, we obtain the explicit expressions for propagators in models with multi-well potentials. In Section 3 we use the obtained potentials and propagators for the non-perturbative analysis of the dynamics of localized states in both symmetric and asymmetric potentials, and for different types of wave packets. In Section 4 we discuss some problems related to the considered problems and future developments.

\section{$N=4~SUSY~QM$ and multi-well potentials}

To construct multi-well potentials in $N=4~SUSY~QM$ \cite{susy-article-ref:12} we add states with energy $\varepsilon $ below the ground state energy $E_{0} $ of the initial Hamiltonian $H_0$ (we assume that $H_0$ possesses only the discrete spectrum). Multi-well structure of the obtained potentials becomes more striking when ${(E_{0} - \varepsilon )\over E_{0} } \ll 1$. The super--Hamiltonian of $N=4~SUSY~QM$ consists of three non-trivial Hamiltonians \cite{susy-article-ref:12} $H_{+}^{-} =H_{-}^{+} =H_{0} -\varepsilon $, $H_{-}^{-} $ and $H_{+}^{+} $. Spectra of the latter two Hamiltonians have the additional state below the ground state of the initial Hamiltonian, when $N=4~SUSY$ is exact, while other states coincide with those of $H_{0} -\varepsilon $. $H_{-}^{-} $ and its wave functions are related to $H_{+}^{-} =H_{0} -\varepsilon $ and its initial wave functions in the following way:
\begin{equation}
  \begin{array}{c}
      \displaystyle H_{-}^{-} =H_{+}^{-} -\frac{d^{2} }{dx^{2} } \ln (\varphi _{1} (x,\varepsilon )+\varphi _{2} (x,\varepsilon )),\\
      \displaystyle \psi _{-}^{-} (x,E)=\frac{1}{\sqrt{2(E_{i} -\varepsilon )} } \frac{W\{ \psi _{+}^{-} (x,E_{i} ),\varphi (x,\varepsilon ,1)\} }{\varphi (x,\varepsilon ,1)},\\
      \displaystyle \psi _{-}^{-} (x,E=0)=\frac{N^{-1} }{\varphi (x,\varepsilon ,1)},~\varphi (x,\varepsilon ,1)=\varphi _{1} (x,\varepsilon )+\varphi _{2} (x,\varepsilon )\\
      \displaystyle N^{-2} =\frac{2W\{ \varphi _{1} ,\varphi _{2} \} }{\Delta (+\infty ,\varepsilon )-\Delta (-\infty ,\varepsilon )},~\Delta (x,\varepsilon )=\frac{\varphi _{1} (x,\varepsilon )-\varphi _{2} (x,\varepsilon )}{\varphi _{1} (x,\varepsilon )+\varphi _{2} (x,\varepsilon )}
  \end{array}
\label{susy-article-eqs:1}
\end{equation}
Here $\varphi _{i} (x,\varepsilon ),~i=1,2$ are two linear independent solutions to the auxiliary equation $H_{0} \varphi (x)=\varepsilon \varphi (x)$.  They are non--negative and have the following asymptotic behavior:  $\varphi _{1} (x)\to +\infty$ ($ \varphi _{2} (x)\to 0$) under $x\to -\infty$, and $\varphi _{1} (x)\to 0$ ($\varphi _{2} (x)\to +\infty )$) under $x\to +\infty$. Here $\psi _{+}^{-} (x,E_{i} )$ are normalized wave functions of $H_{0} $, $W\{ \varphi _{1} ,\varphi _{2} \} $ is the Wronskian.

Using the form--invariance of $H_{+}^{+}$ and $H_{-}^{-}$ \cite{susy-article-ref:12} one could obtain similar expressions for $H_{+}^{+} $, $\psi _{+}^{+} $:
  \begin{equation}
    \begin{array}{c}
	\displaystyle H_{+}^{+} =H_{+}^{-} -\frac{d^{2} }{dx^{2} } \ln \left(\varphi _{1} (x,\varepsilon )+\Lambda (\varepsilon ,\lambda )\; \varphi _{2} (x,\varepsilon )\right),\\
	\displaystyle \psi _{+}^{+} (x,E=0)=\frac{N_{\Lambda }^{-1} }{\left(\varphi _{1} (x,\varepsilon )+\Lambda (\varepsilon ,\lambda )\; \varphi _{2} (x,\varepsilon )\right)}\\
	\displaystyle \psi _{+}^{+} (x,E_{i} )=\frac{1}{\sqrt{2(E_{i} -\varepsilon )} } \frac{W\{ \psi _{+}^{-} (x,E_{i} ),\varphi _{1} (x,\varepsilon )+\Lambda (\varepsilon ,\lambda )\varphi _{2} (x,\varepsilon )\} }{\left(\varphi _{1} (x,\varepsilon )+\Lambda (\varepsilon ,\lambda )\varphi _{2} (x,\varepsilon )\right)} ,\\
	\displaystyle \Lambda (\varepsilon ,\lambda )=\frac{\Delta (\infty ,\varepsilon )-\lambda -(\lambda +1)\; \Delta (-\infty ,\varepsilon )}{\Delta (\infty ,\varepsilon )+\lambda -(\lambda +1)\; \Delta (-\infty ,\varepsilon )},
    \end{array}
  \label{susy-article-eqs:2}
  \end{equation}
where the parameter $\lambda$ is restricted to be $\lambda >-1$, and the normalization constant is $N_{\Lambda }^{-2} =(1+\lambda )\; N^{-2} $.

In what follows, we will use the Hamiltonian of HO as the initial Hamiltonian to consider the tunneling of wave packets. Thus, the solutions to the auxiliary equation are parabolic cylinder functions $\varphi _{1} (\xi ,\bar{\varepsilon })=D_{\nu } (\sqrt{2} \xi ),~\varphi _{2} (\xi ,\bar{\varepsilon })=D_{\nu } (-\sqrt{2} \xi )\; $ and the Wronskian becomes $W\{ \varphi _{1} ,\varphi _{2} \} =\frac{2\sqrt{\pi \omega } }{\Gamma (-\nu )} $ \cite{susy-article-ref:19}, where $\Gamma (-\nu )$ is the gamma-function, $\xi =\sqrt{\omega \; } x$, $\nu =-\frac{1}{2} +\frac{\varepsilon }{\omega } =-\frac{1}{2} +\bar{\varepsilon }$:
  \begin{equation}
      \varphi (\xi ,\bar{\varepsilon },1)=D_{\nu } (\sqrt{2} \xi )+\; D_{\nu } (-\sqrt{2} \xi ).
      \label{susy-article-eqs:3}
  \end{equation} 

\begin{figure}
	    \begin{tabular}{cc}
	    I \epsfig{file=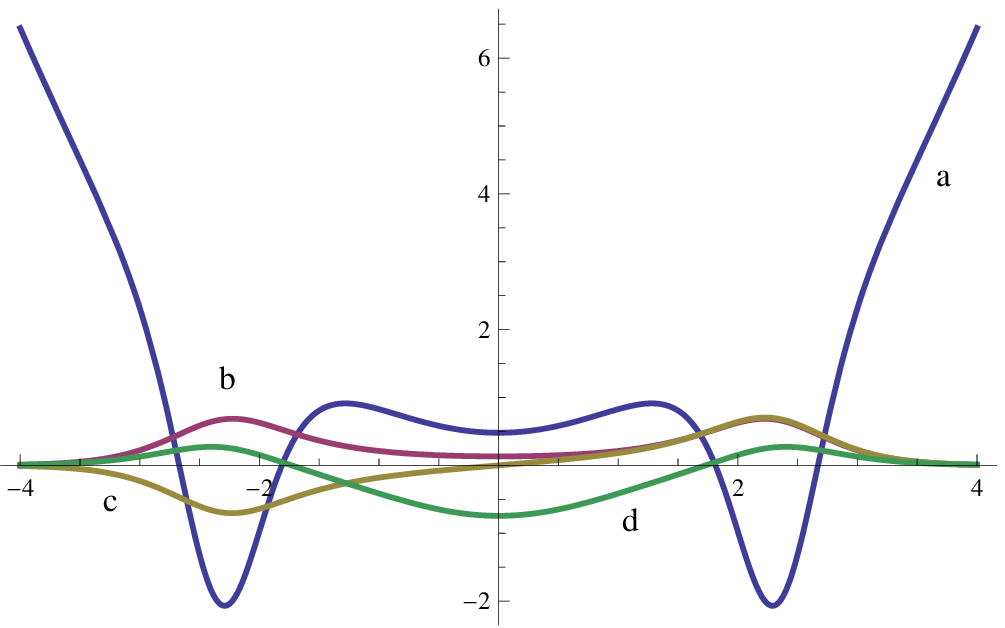, width=0.45\linewidth}
	    &
	    II \epsfig{file=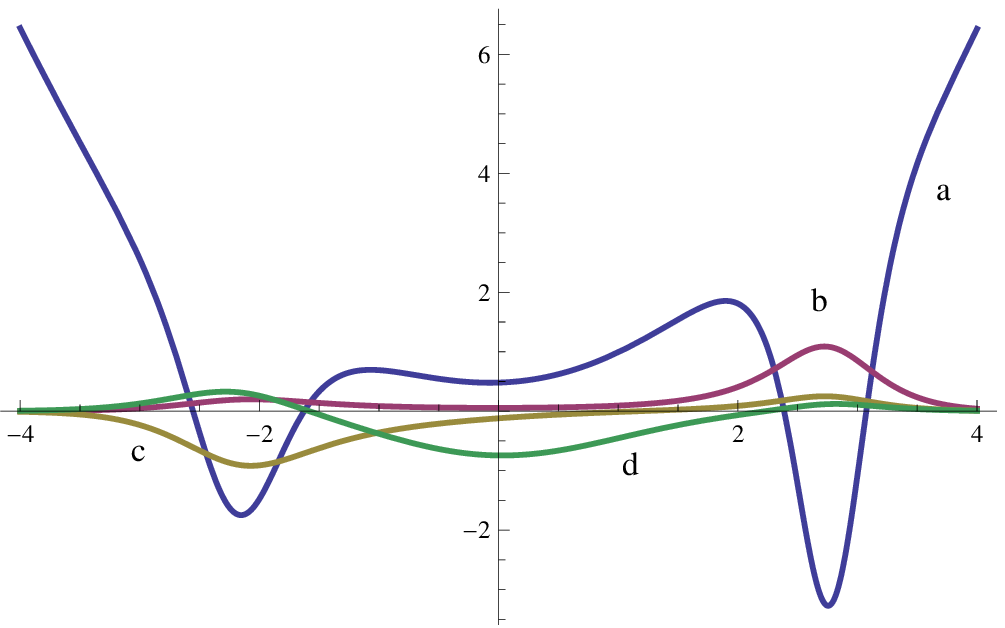, width=0.45\linewidth}
	    \end{tabular}
	    \caption{Potentials and wave functions of Hamiltonians $H_-^-$ and $H_+^+$. Left: (a) -- potential $U_{-}^{-}(\xi)$ ($\omega = 1$, $\nu=-0.02$), wave functions of (b) -- ground state, (c) -- first excited and (d) -- second excited states (d). Right: (a) -- potential $U_ + ^ + (\xi ,\lambda+1 )$ ($\omega = 1$, $\nu=-0.02$, $\lambda=-0.95$) and corresponding wave functions.}
	    \label{susy-article-fig:1}
     \end{figure}

Note that $\Delta (\pm \infty ,\varepsilon ,\lambda )$, entering $\Lambda (\varepsilon ,\lambda )$, are determined by the asymptotes of the solutions $\varphi _{1} (\xi ,\bar{\varepsilon }),~\varphi _{2} (\xi ,\bar{\varepsilon })$. The ground state wave function of $H_{-}^{-} $ is
  \begin{equation}
      \psi _{-}^{-} (x,E=0)=\frac{N^{-1} }{\left(D_{\nu } (\sqrt{2} \xi )+D_{\nu } (-\sqrt{2} \xi )\right)} =\frac{N^{-1} }{\varphi (\xi ,\bar{\varepsilon },1)} ,\quad \; N^{-2} =\frac{2\sqrt{\pi \omega } }{\Gamma (-\nu )}  ,
      \label{susy-article-eqs:4}
  \end{equation} 
and wave functions of excited states are determined by (\ref{susy-article-eqs:1}), where $\psi _{+}^{-} (x,E_{n} )=\left(\frac{\omega }{\pi }\right)^{1\over 4} \frac{1}{\sqrt{n!} } D_{n} (\sqrt{2} \xi )$, $n=0,1,2\ldots$ are the wave functions of HO. The potential of the Hamiltonian $H_{-}^{-} (p,x)=\omega \left(-\frac{1}{2} \frac{d^{2} }{d\xi ^{2} } +(\frac{\xi ^{2} }{2} -\bar{\varepsilon })-\frac{d^{2} }{d\xi ^{2} } \ln \varphi (\xi ,\bar{\varepsilon },1)\right)$ (see Fig. \ref{susy-article-fig:1}) is defined  by the symmetric combination of solutions $\varphi _{1} (x,\varepsilon )+\varphi _{2} (x,\varepsilon )$, while the potential of $H_{+}^{+} $ is defined by the asymmetric combination $\varphi _{1} (x,\varepsilon )+(\lambda +1)\; \varphi _{2} (x,\varepsilon )$, and corresponds to the family of Hamiltonians with different values of $\omega $. Hence, the ground state wave function of $H_{+}^{+} $ has the form of
  \begin{equation}
      \psi _{+}^{+} (x,E=0)=\frac{N_{\lambda } ^{-1} }{\left(D_{\nu } (\sqrt{2} \xi )+(\lambda +1)D_{\nu } (-\sqrt{2} \xi )\right)} =\frac{N_{\lambda } ^{-1} }{\varphi (\xi ,\bar{\varepsilon },\lambda +1)} ,\; \; N_{\lambda } ^{-2} =\frac{2(\lambda +1)\sqrt{\pi \omega } }{\Gamma (-\nu )}.
      \label{susy-article-eqs:5}
  \end{equation}
      


It should be noted that in terms of a dimensionless variable $\xi $ the only way to vary the form of the potential is to vary $0<\bar{\varepsilon }<1/2$ and $-1<\lambda $. In the case of natural units $x$, additionally, the form of the potential (in particular, the position of local minima) can be changed by variation of $\omega $.

\section{Time evolution of states in $N=4~SUSY~QM$}

In many cases tunneling is considered within the two--mode approximation. It allows to describe density oscillations  and revival times for wave packets, but this approach fails, for example, in reproducing the dependence of the wave packet dynamics on the shape of the potential and the coherence breaking \cite{susy-article-ref:14}. For instance, to correctly describe the dynamics of squeezed wave packets (WP) initially localized in one of the minima,  one has to take into account not only the ground and first excited states, but also  higher excited states. Contribution of these states increases with the squeezing of WP, and is also significant in potentials with low barriers, as for example in SQUIDs (superconducting quantum interference devices) and in cold atoms traps.

The time evolution of the Gaussian wave packet $\Phi (x)=\left(\frac{\omega e^{2R} }{\pi } \right)^{1/4} \exp \left(-\frac{1}{2} \left(x-x_{0} \right)^{2} e^{2R} \right)$ (here $R$ is the squeezing parameter) initially localized in $x=x_{0} $ is determined by \cite{susy-article-ref:20}
  \begin{equation} \label{susy-article-eqs:6}
      \Phi \left(x,t\right)=\int _{-\infty }^{+\infty }K\left(x,t;x_{0} ,0\right) \; \Phi (x_{0} )\; dx_{0} ,~K\left(x,t;x_{0} ,0\right)=\sum _{n=0}^{\infty }\psi _{n}  (x)\; \psi _{n}^{*} (x_{0} )\; e^{-iE_{n} t}.
      \label{susy-article-eqs:6}
  \end{equation} 
Here $K\left(x,t; x_{0} ,0\right)$ is the propagator, which sufficiently describes the dynamics of localized states in potentials of arbitrary complexity.

Currently only a few exactly-solvable models with analytic expressions of the propagator $K\left(x,t; x_{0} ,0\right)$ are known \cite{susy-article-ref:21}. Usually, their Hamiltonians have one-well potentials. The construction of new models and their propagators on the basis of propagators of exactly-solvable models with one--well potentials is proposed in  \cite{susy-article-ref:10, susy-article-ref:17, susy-article-ref:18}. Main ideas of this approach are outlined in  \cite{susy-article-ref:10, susy-article-ref:17, susy-article-ref:18}, so we will briefly discuss the procedure of the construction of propagators in  $N=4~SUSY~QM$, starting from exactly solvable model with the confinement potential. Let's denote propagators, corresponding to Hamiltonians $H_{\sigma _{1} }^{\sigma _{2} } $ of $N=4~SUSY~QM$, as $K_{\sigma _{1} }^{\sigma _{2} } \left(x,t;x_{0} ,0\right)\, (\sigma _{i} =\pm )$. Using the form-invariance of $N=4~SUSY~QM$ potentials, established in \cite{susy-article-ref:12}, the expression for the propagator $K_{+}^{+} \left(x,t;x_{0} ,0\right)$ could be obtained: $K_{+}^{+} \left(x,t;x_{0} ,0\right)$ is related to the initial $K_{+}^{-} \left(x,t;x_{0} ,0\right)$ of the exactly-solvable model as follows
  \begin{equation}
      K_{+}^{+} \left(x,t; y,0\right)=\frac{1}{2} L_{x} L_{y} \int _{-\infty }^{+\infty }dzK_{+}^{-} \left(x,t; z,0\right)\;  G_{+}^{-} (z,y,\varepsilon )\; +\frac{N_{\Lambda }^{-2} \; e^{-i\varepsilon t} }{\varphi (x,\varepsilon ,\Lambda )\; \varphi (y,\varepsilon ,\Lambda )}  .
   \label{susy-article-eqs:7}
  \end{equation} 
Here $L_{x} =\left(\frac{d}{dx} -\frac{\varphi '(x,\varepsilon ,\Lambda )}{\varphi (x,\varepsilon ,\Lambda )} \right)$ and $G_{+}^{-} (z,y,\varepsilon )$ is the Green function of the Schr\"{o}dinger equation with energy $\varepsilon $:
\[G_{+}^{-} (x,y,\varepsilon )=-\frac{2}{W\{ f_{l} ,f_{r} \} } \left(f_{l} (x,\varepsilon )\; f_{r} (y,\varepsilon )\theta (y-x)\; +f_{l} (y,\varepsilon )\; f_{r} (x,\varepsilon )\theta (x-y)\right).\] 
According to our notation, $f_{l} (x,\varepsilon )=\varphi _{2} (x,\varepsilon ),\; f_{r} (x,\varepsilon )=\varphi _{1} (x,\varepsilon )$. Acting with the operator $L_{y} $ and simplifying the expression, the  propagator becomes
  \begin{equation}
      \begin{array}{c}
	  \displaystyle K_{+}^{+} \left(x,t; y,0\right)=-\frac{1}{\varphi (y,\varepsilon ,\Lambda )} \, L_{x} \left[\Lambda (\varepsilon ,\lambda )\, \int _{-\infty }^{y}dz\, K_{+}^{-} \left(x,t; z,0\right)\, \varphi _{2} (z,\varepsilon )-\right.\\
	  \displaystyle\left.\int _{y}^{\infty }dz\, K_{+}^{-} \left(x,t; z,0\right)\, \varphi _{1} (z,\varepsilon )  \right]\, +\frac{N_{\Lambda }^{-2} \; e^{-i\varepsilon t} }{\varphi (x,\varepsilon ,\Lambda )\varphi (y,\varepsilon ,\Lambda )}.
      \end{array}
    \label{susy-article-eqs:8}
  \end{equation}
Choosing the Hamiltonian of HO as $H_{+}^{-} $, we get $\varphi _{1} (\xi ,\bar{\varepsilon })=D_{\nu } (\sqrt{2} \xi ),~\varphi _{2} (\xi ,\bar{\varepsilon })=D_{\nu } (-\sqrt{2} \xi )$, $\Lambda \left(\varepsilon ,\lambda \right)=\left(1+\lambda \right)$, and $K_{+}^{-} \left(x,t;y,0\right)$ becomes \cite{susy-article-ref:21}:
  \begin{equation}
      K_{+}^{-} \left(x,t;\, y,0\right)=\left(\frac{\omega \, e^{-i\pi (\frac{1}{2} +n)} }{2\pi \, \sin \omega \tau } \right)^{1/2} \exp \left\{\frac{i\omega }{2\, \sin \omega \; t} \left[\left(x^{2} +y^{2} \right)\, \cos \omega \; t-2xy\right]\, \right\},
    \label{susy-article-eqs:9}
  \end{equation} 
where $t={n\pi\over \omega  } +\tau,~n \in N_{0}, 0<\tau <{\pi \over\omega  }$. The factor $e^{-i\pi (\frac{1}{2} +n)}$ ensures correct behavior of the propagator for all values of time $t$ and sewing of at $\omega t = \pi n$. Expressions (\ref{susy-article-eqs:4}),  (\ref{susy-article-eqs:5}), (\ref{susy-article-eqs:6}), (\ref{susy-article-eqs:8}) and (\ref{susy-article-eqs:9}) are the basic expressions to study the dynamics of localized states both in symmetric and asymmetric potentials. At the same time, this approach takes into account all the states, which form the localized state $\Phi (x,0)$. $K_{-}^{-} \left(x,t;\, y,0\right)$ could be obtained from (\ref{susy-article-eqs:8}) by substitution $\Lambda \left(\varepsilon ,\lambda \right)=1$ and corresponds to a symmetric potential.

\section{Dynamics of localized states in multi-well potentials}

In this paper we focus on the case, when only a few states of the Hamiltonian with the multi--well potential are located below the barrier. This is a common situation for different physical problems, both in atomic and in solid state physics. In general, the dynamics of the localized states can not be correctly described in the traditional framework  of the tunnel splitting $\Delta =E_{1} -E_{0} $, because the initially localized state can not be expanded as a superposition of the wave functions of the tunnel duplet. Higher excited states are essential in $\Phi (x,0)$ and their contribution increases with increasing the localization of the wave packet. In contrast, the approach, considered in Section 3, allows one to study the dynamics of localized states taking into account all states of the exactly-solvable Hamiltonian, both with symmetric and asymmetric multi-well potentials. Since the spectra of $H_{-}^{-} $ and $H_{+}^{+} $ are identical, the tunnel splittings $\Delta =E_{1} -E_{0} $ are equal for symmetric and asymmetric potentials.

 We will study the dynamics of the localized states in potentials constructed from the potential of HO (\ref{susy-article-eqs:1})-(\ref{susy-article-eqs:5}), with the propagator (\ref{susy-article-eqs:8}), (\ref{susy-article-eqs:9}). Their expressions in the dimensionless variables are
  \begin{equation}
  \begin{array}{c}
     \displaystyle K_{+}^{+} \left(\xi,T;\eta ,0\right)=-\frac{\sqrt{\omega } }{\varphi (\eta ,\bar{\varepsilon },\lambda +1)} \, L_{\xi } \left[(\lambda +1)\, \int _{-\infty }^{\eta }d\zeta \, K_{+}^{-} \left(\xi ,T;\; \zeta ,0\right)D_{\nu } (-\sqrt{2\zeta } )\right.- \\
     \displaystyle \left.-\int _{\eta }^{\infty }d\zeta \, K_{+}^{-} \left(\xi ,T;\; \zeta ,0\right)\, D_{\nu } (\sqrt{2\zeta } )  \right] +\frac{(\lambda +1)N^{-2} \; e^{-i\bar{\varepsilon }\; T} }{\varphi (\xi ,\bar{\varepsilon },\lambda +1)\; \varphi (\eta ,\bar{\varepsilon },\lambda +1)} ,\; \; \; \bar{\varepsilon }=\frac{\varepsilon }{\omega },\\
      \displaystyle K_{+}^{-} \left(\xi ,T;\eta ,0\right)=\left(\frac{\, e^{-i\pi (\frac{1}{2} +n)} }{2\pi \, \sin \tau } \right)^{1/2} \exp \left\{\frac{i}{2\, \sin T} \left[\left(\xi ^{2} +\eta ^{2} \right)\, \cos T-2\xi \eta \right]\, \right\},
  \end{array}
    \label{susy-article-eqs:11}
  \end{equation}
where $T=\pi n+\tau ,\; n\in N,\; 0<\tau <\pi$. The propagator $K_{-}^{-} \left(\xi ,T;\; \eta ,0\right)$ describing the dynamics of wave packets in the symmetric potential can be obtained from (\ref{susy-article-eqs:11}) with $\lambda =0$. Thus, the expression (\ref{susy-article-eqs:6}) in the dimensionless variables reads
  \begin{equation}
      \Phi \left(\xi ,\; T\right)=\int _{-\infty }^{+\infty }d\zeta \; K_{+}^{+}  \left(\xi ,T;\zeta ,0\right)\; \Phi \left(\zeta ,\; 0\right)\; ,\quad \Phi \left(\zeta ,\; 0\right)=\left(\frac{e^{2R} }{\pi } \right)^{1/4} e^{-\frac{\zeta ^{2} }{2\sigma^2} },~\sigma^2=e^{-2R}.
    \label{susy-article-eqs:12}
  \end{equation}
These relations allow one to obtain the form of the wave packet as a function of time and spatial coordinates.
\newpage

 \textbf{Small squeezed states.}

\begin{figure}[h]
	    \begin{tabular}{cc}
	    a \epsfig{file=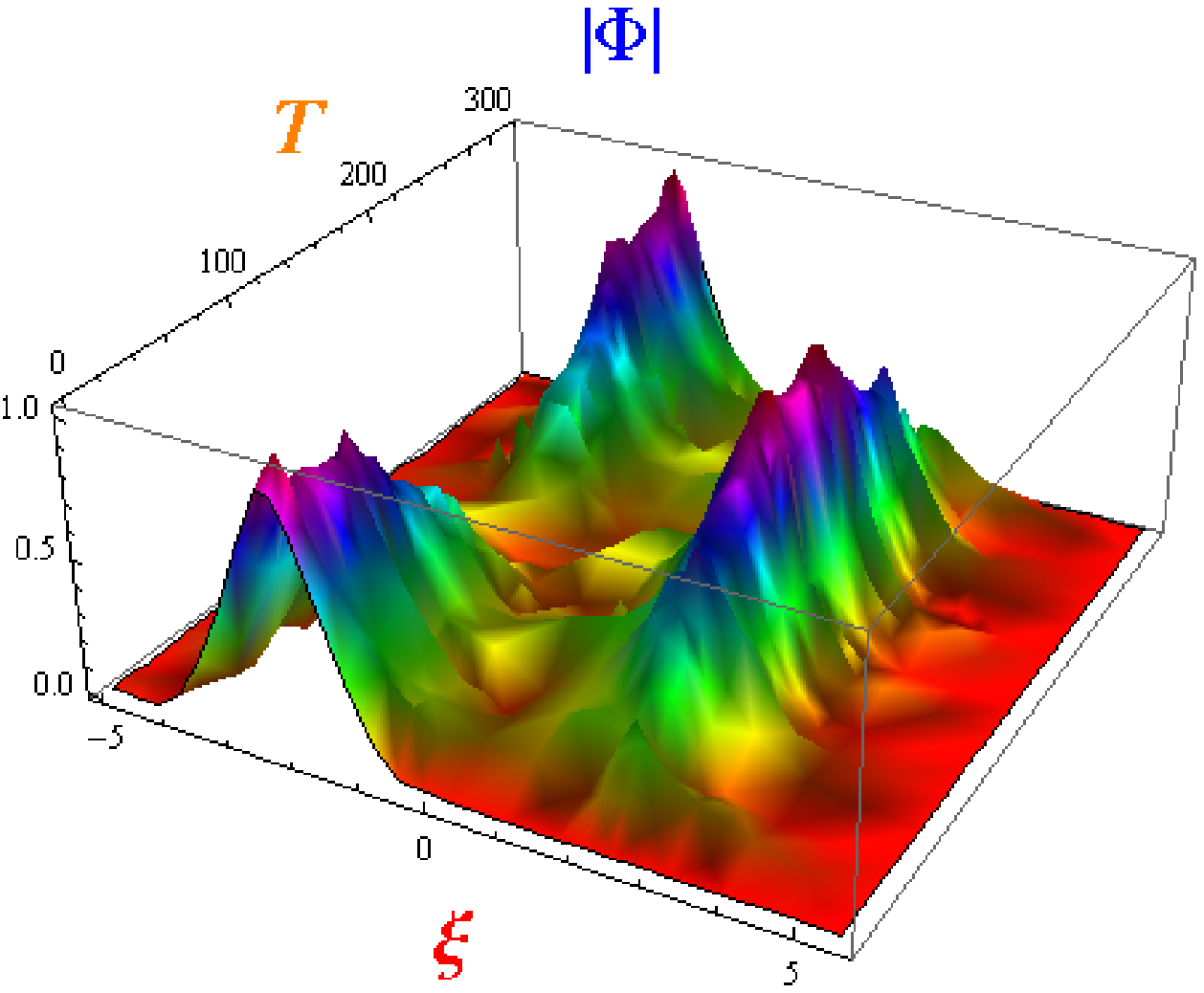, width=0.45\linewidth}
	    &
	    b \epsfig{file=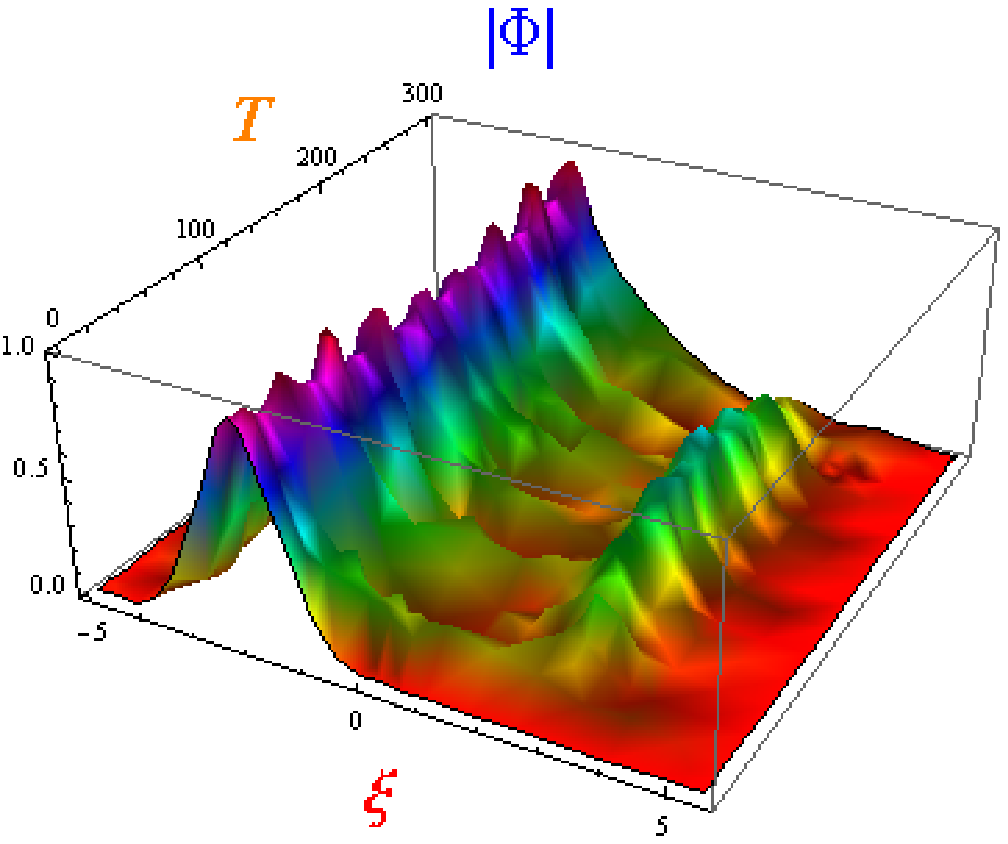, width=0.45\linewidth}
	    \end{tabular}
	    \caption{Time dependence of wave packet $\left|\Phi (\xi ,T)\right|$ with $R=0.35$, initially localized in the left minimum of (a) -- potential $U_{-}^{-} \left(\xi \right)$ $\left(\xi _{l} =-2.29\right)$, (b) -- potential $U_{+}^{+} \left(\xi ,\lambda \right)$ with $\lambda =-0.95$ $(\xi _{l} =-2.153)$.}
	    \label{susy-article-fig:2}
     \end{figure}

 Let's note that study of the dynamics of the localized states (\ref{susy-article-eqs:12}) can not be performed in the two-mode approximation, even when $R=0$. A satisfactory approximation of the initially localized state $\Phi (\xi ,0)$ is achieved with taking into account eight states of the Hamiltonian $H_{-}^{-} $ (see line 1 in table 1). At the same time, the expansion of the initial state $\Phi \left(\xi ,0\right)$ over states of $H_{-}^{-} \left(H_{+}^{+} \right)$ is not required to compute $\left|\Phi \left(\xi ,T\right)\right|$ when using the exact propagator. We will compare the results of the exact calculation using (\ref{susy-article-eqs:12}) with $\left|\Phi (\xi ,T)\right|=\left|\sum _{n=0}^{n_{\max } }c_{n} \, \psi _{\left(-\right)+}^{\left(-\right)+} (\xi ,E_{n} )\exp (-iE_{n} T) \right|$ to demonstrate the effectiveness of the basis of $H_{-}^{-} \left(H_{+}^{+} \right)$ in the considered problem. Fig.\ref{susy-article-fig:2} shows $\left|\Phi \left(\xi ,T\right)\right|$ for the same, as on Fig.\ref{susy-article-fig:1}, potentials $U_{-}^{-} (\xi )$ and $U_{+}^{+} (\xi ,\lambda )$, and for the value of squeezing parameter $R=0.35$. The latter corresponds to the weak localization. Initially (at $T=0$) the wave packet is localized in the left local minimum and has the energy  ($E_{\Phi } ={\left\langle \Phi  \right|} H_{-}^{-} (p,x)\, {\left| \Phi  \right\rangle} $) $E_{\Phi } =0.16$ for the symmetric potential, and $E_{\Phi } =0.185$ for the asymmetric one.

\begin{table}[h]
\begin{footnotesize}
\begin{tabular}{|l||p{0.35in}|p{0.4in}|p{0.35in}|p{0.4in}|p{0.4in}|p{0.4in}|p{0.35in}|p{0.4in}|p{0.35in}|p{0.4in}|} \hline
\# of state & 0 & 1 & 2 & 3 & 4 & 5 & 6 & 7 & 8 & 9 \\ \hline \hline
$\lambda =0$, $R=0$ & 0.668 & -0.664 & 0.018 & 0.017 & 0.048 & -0.146 & 0.203 & -0.184 & 0.110 & -0.031 \\ \hline
$\lambda =0$, $R=0.35$ & 0.682 & -0.692 & 0.135 & -0.082 & 0.056 & -0.069 & 0.094 & -0.090 & 0.044 & 0.015 \\ \hline
$\lambda =-0.95$, l & 0.208 & -0.945 & 0.179 & -0.105 & 0.065 & -0.066 & 0.077 & -0.060 & 0.013 & 0.029 \\ \hline
$\lambda =-0.95$, r & 0.941 & 0.213 & 0.038 & 0.011 & -0.007 & 0.011 & 0.066 & 0.128 & 0.153 & 0.115 \\ \hline
\end{tabular}
\end{footnotesize}
\caption{Coefficients of the expansion of the initial wave packet (\ref{susy-article-eqs:12}).}
\label{susy-article-tbl:1}
\end{table}

 Approximation of the initial WP by states of $H_{-}^{-} $ and $H_{+}^{+} $ is mainly determined by few lower states (see Table \ref{susy-article-tbl:1}) which indicate the suitability of the basis. Nevertheless, despite a small contribution of higher states, they determine fine details of time evolution of WP, as we will show later. These details include specific beats caused by the interference of excited states. When the asymmetric potential is used, the expansion of the initial packet contains more terms, than in the symmetric potential (see Table \ref{susy-article-tbl:1}), and at the same time the contribution of states with $n>9$ is comparable to the contribution of low-lying states.

Thus, the temporal dynamics of WP (Fig. \ref{susy-article-fig:2}a) demonstrates slow tunnel transition of under-barrier states and fast oscillations of over-barrier states, which have higher intensity in compare to the symmetric case. In the case of a symmetric potential the evolution of $\left|\Phi \left(\xi ,T\right)\right|$ (Fig. 2a) has the striking oscillatory nature: the portion of the initial $\Phi \left(\xi ,0\right)$ which tunnels to the right minimum of $U_{-}^{-} (\xi )$ is quite large and reaches its maximum at $T\simeq \frac{T_{rev} }{2} $ ($T_{rev} =\frac{2\pi }{(E_{1} -E_{0} )} \approx 300$), while at $T\simeq T_{rev} $ the WP is completely restored at the left minimum of $U_{-}^{-} (\xi )$. At the same time, the contribution of higher excited states to $\left|\Phi \left(\xi ,T\right)\right|$  is relatively small and leads only to small ''beats''. Completely different dynamics is observed, when the WP is initially localized in one of the minima of $U_{+}^{+} (\xi ,\lambda )$ (Fig. \ref{susy-article-fig:2}b). The portion of the WP which tunnels to another local minimum is small, since in the left local minimum of $U_{+}^{+} (\xi ,\lambda )$ (at $\xi _{l} =-2.153$)  the largest contribution to $\Phi \left(\xi ,0\right)$ comes from the first excited state of $H_{+}^{+} $ with the wave function being very small in the right well. It means, that this state contributes only a small portion to the tunnel transition amplitude. The similar situation is observed, when the wave packet is localized at the right local minimum $(\xi _{r} =2.755)$ of $U_{+}^{+} (\xi ,\lambda )$. In this case the ground state wave function of $H_{+}^{+} $, which is small in the left well, contributes mostly to $\Phi \left(\xi ,0\right)$. In some sense the wave packet is trapped within the initial well. The mechanism of such partial trapping of WP is simple:  if one of the under-barrier states mostly contributes to $\Phi \left(\xi ,0\right)$, then its wave function is small in another well, that means that it has a small contribution to the tunnel transition to another well. Other under--barrier states give small contributions to tunneling due to their small portion in the formation of $\Phi \left(\xi ,0\right)$. Nevertheless, the contribution of over--barrier states is larger than that in the symmetric potential, and it leads to large beats.

 The phenomenon of partial trapping is more obvious when the initial wave packet is uniformly distributed among the local minima $\left(\xi _{L} \, ,\; \xi _{R} \right)$ of the asymmetric potential $U_{+}^{+} \left(\xi ,\lambda \right)$, e.g.
  \begin{equation}
      \Phi (\xi ,0)=\left(\frac{\; \sigma ^{-2} }{4\pi } \right)^{1/4} \; \frac{e^{-\frac{(\xi -\xi _{L} )^{2} }{2\sigma ^{2} } } +e^{-\frac{(\xi -\xi _{R} )^{2} }{2\sigma ^{2} } } }{\sqrt{\left(1+e^{-\, \frac{1}{4\sigma ^{2} } (\xi _{L} -\xi _{R} )^{2} } \right)} } ,\; \; \sigma ^{2} =e^{-2R} .
    \label{susy-article-eqs:13}
  \end{equation} 
The dynamics of tunneling of the initial $\Phi \left(\xi ,0\right)$ (\ref{susy-article-eqs:13}) differs in different local minima of $U_{+}^{+} \left(\xi ,\lambda \right)$ (Fig. 3).



\begin{figure}
  \begin{center}
  \includegraphics[scale=1]{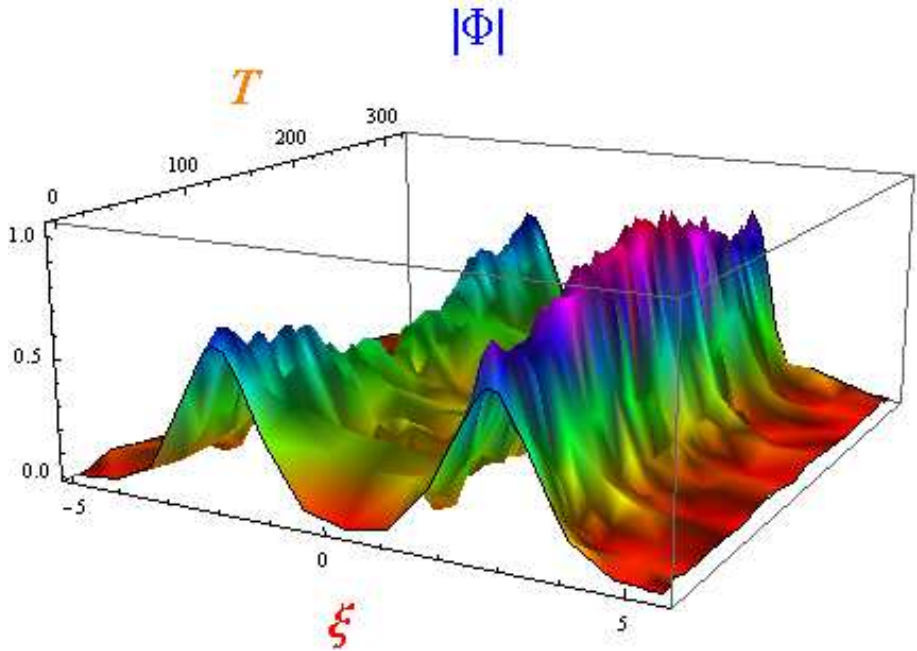}
  \end{center}
  \caption{Time dependence of $\left|\Phi (\xi ,T)\right|$ for distributed initial packet (\ref{susy-article-eqs:13}) in potential $U_{+}^{+} \left(\xi ,\, \lambda \right)$ with $\lambda =-0.95$.}
 \label{susy-article-fig:3}
\end{figure}



\begin{figure}
	    \begin{tabular}{cc}
	    a \epsfig{file=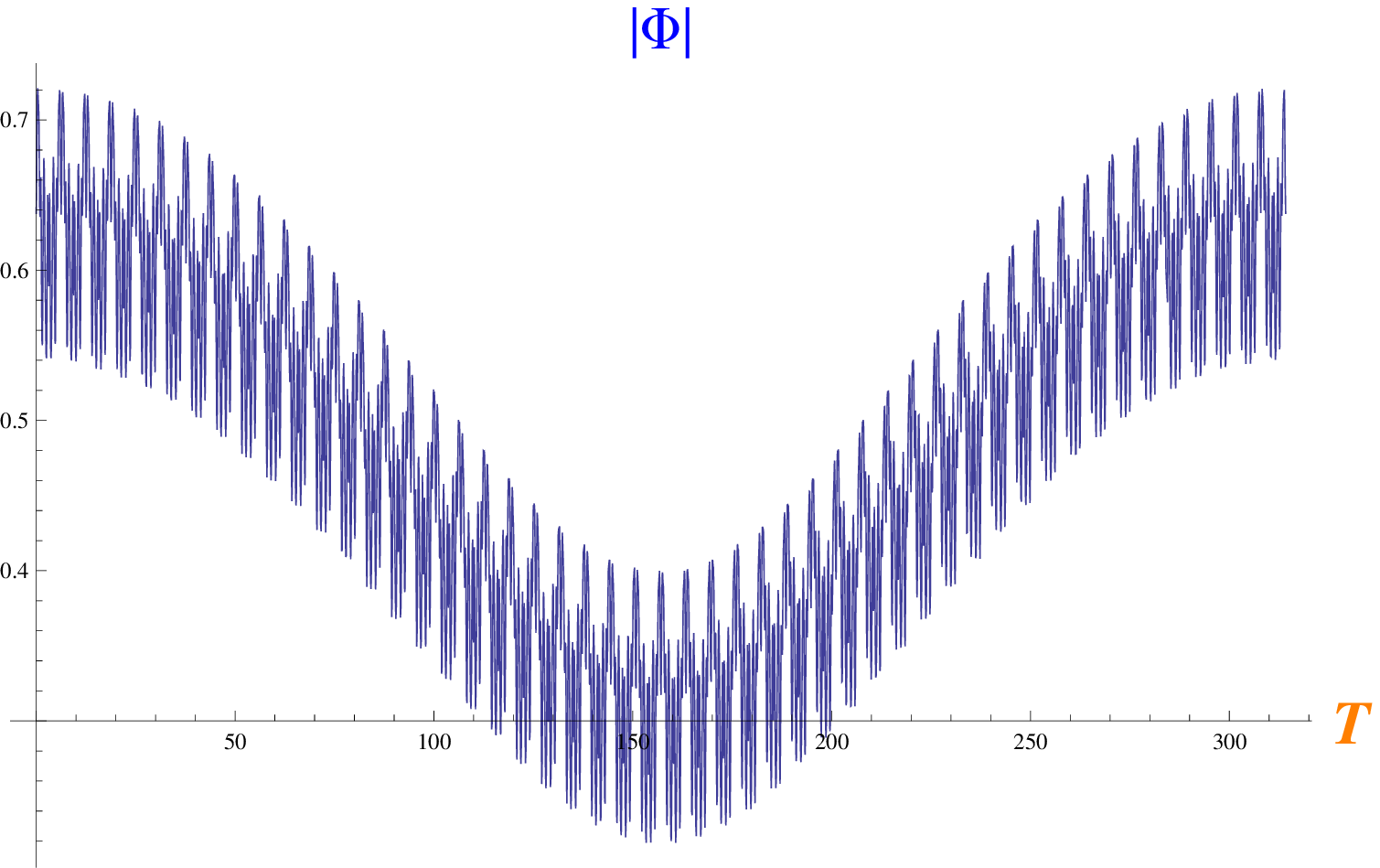, width=0.45\linewidth}
	    &
	    b \epsfig{file=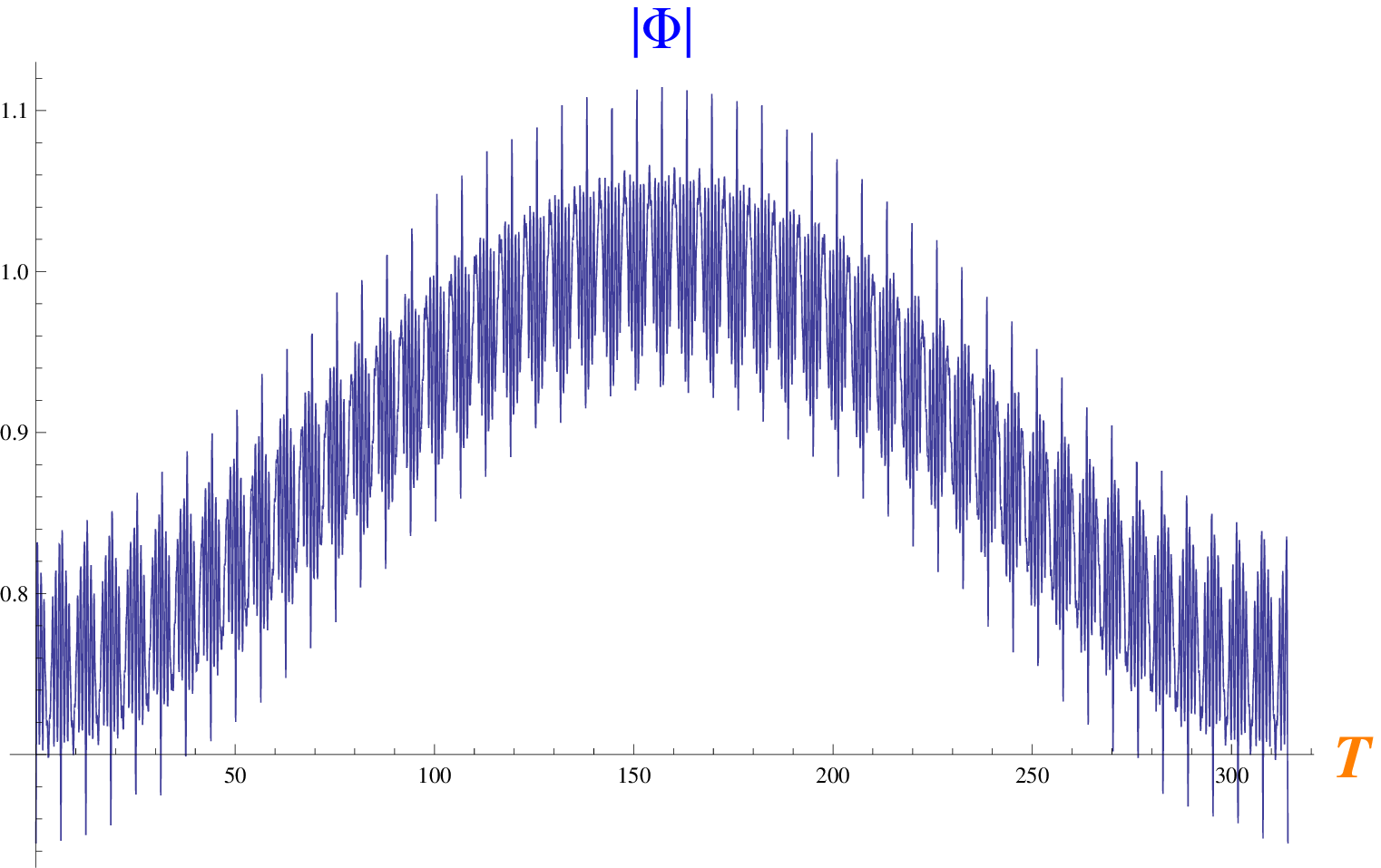, width=0.45\linewidth}
	    \end{tabular}
	    \caption{Time dependence of $\left|\Phi (\xi ,T)\right|$ with $R=0.35$ in (a) -- left well $(\xi _{l} =-2.153)$, (b) -- right well $(\xi _{r} =2.755)$.}
	    \label{susy-article-fig:4}
     \end{figure}

 Fig. \ref{susy-article-fig:3} shows that the portion of WP (\ref{susy-article-eqs:13}) behaves differently in different wells of $U_{+}^{+} \left(\xi ,\lambda \right)$. In the left minimum, for example, $\left|\Phi \left(\xi ,T\right)\right|$ clearly oscillates, but at the same time the dynamics of $\left|\Phi \left(\xi ,T\right)\right|$ in the right minimum is more complicated. To analyze it we give the time dependence of the square of the wave packet in minima of the left $(\xi _{l} =-2.153)$ and the right $(\xi _{r} =2.755)$ wells (Fig. \ref{susy-article-fig:4}).

 It's important to note that in the left well wave packet oscillates and is completely restored after the time interval $T_{rev} =\frac{2\pi }{(E_{1} -E_{0} )} \approx 300$. At the same time, the fraction of the wave packet in the right well increases at $0<T<150$ while $\left|\Phi (\xi ,T)\right|$ is completely changed due to the tunneling from the left well. Moreover, when the value of time is close to $T\sim 150$, the strong squeezing of the packet occurs, and under increasing of $T$  $\left|\Phi (\xi ,T)\right|$ decreases and reaches its initial value. It looks like  partial ''confining'' of the portion of WP inside the right well. Thus, in one of the minima (the left one) the tunneling dynamics possesses the oscillatory nature while in another minimum the partial ''trapping'' of a part of WP occurs.



\begin{figure}

	    \begin{tabular}{cc}
	    a \epsfig{file=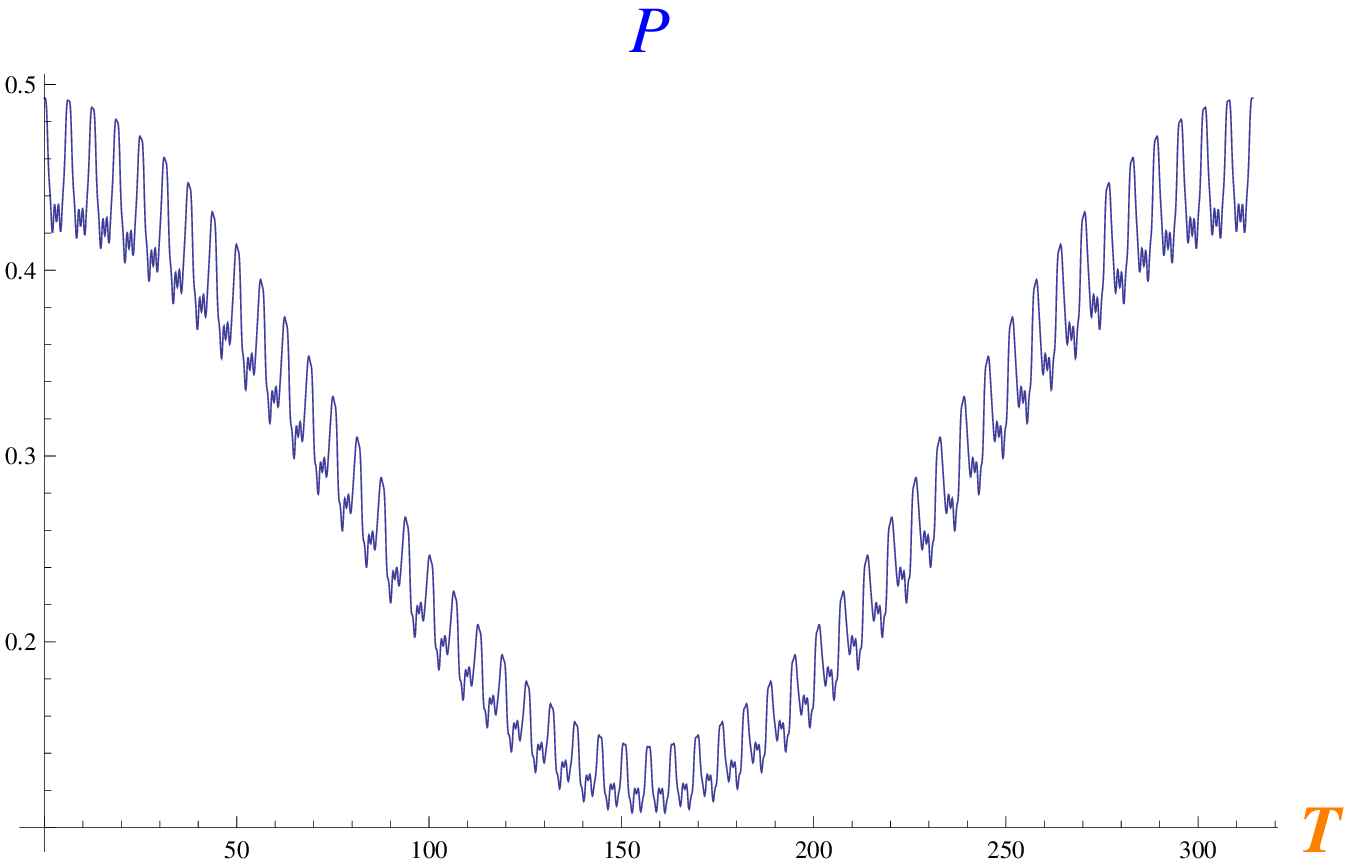, width=0.45\linewidth}
	    &
	    b \epsfig{file=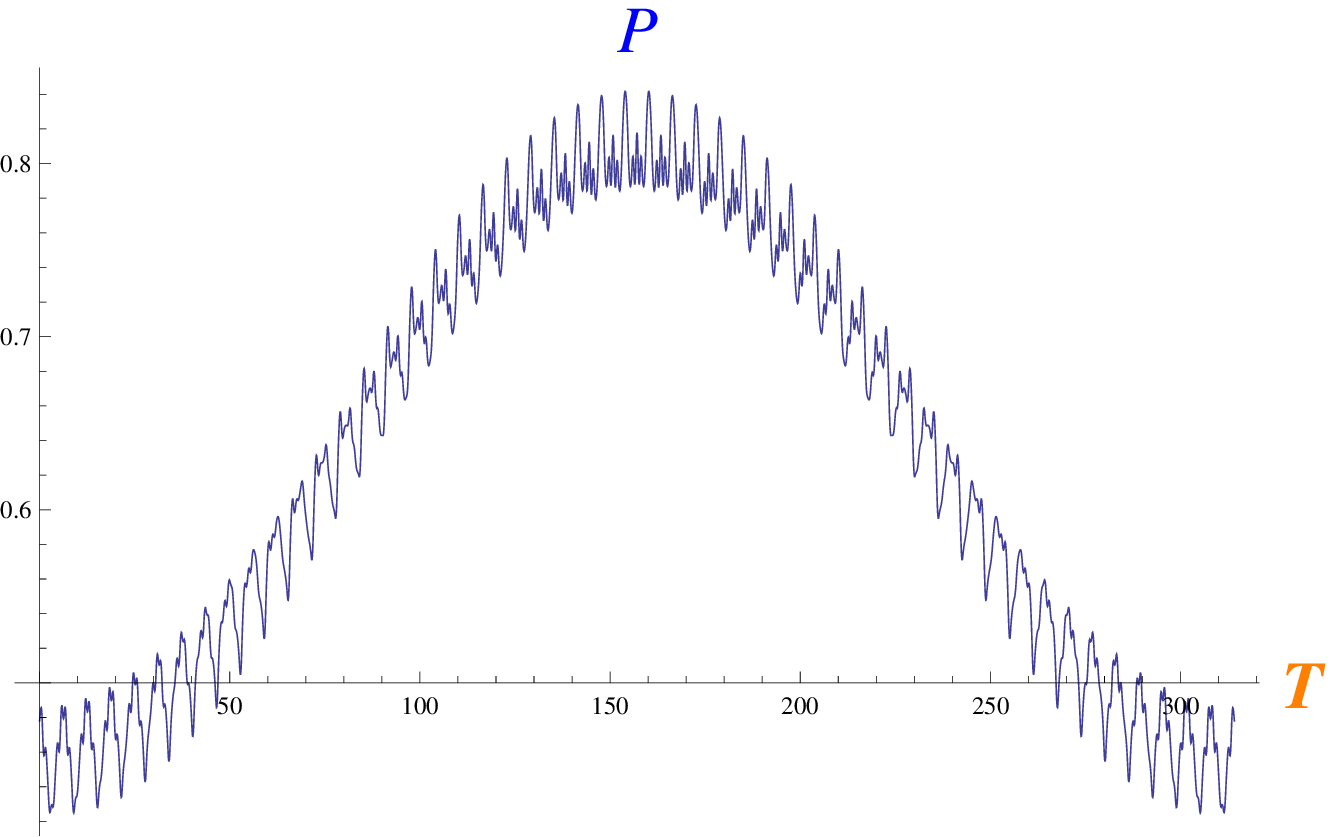, width=0.45\linewidth}
	    \end{tabular}
	    \caption{$P_{l(r)} (T)$ for the initial packet $\Phi (\xi ,0)$ (\ref{susy-article-eqs:13}) with $R=0.35$.}
	    \label{susy-article-fig:5}
     \end{figure}

Another important characteristic of the WP tunneling dynamics is the probability to find wave packet in a certain well, $P_{l(r)} (T)=\int _{x\in l(r)well}\left|\Phi (x,T)\right| ^{2} dx$. Fig. \ref{susy-article-fig:5} shows $P_{l(r)} (T)$ for the potential $U_{+}^{+} (\xi ,\lambda )$ and the initial localized state (\ref{susy-article-eqs:13}). $P_{l(r)} (T)$ is a quantitative characteristics of the localized state dynamics, and it can be revealed that a packet with the same distribution evolves in different local minima of $U_{+}^{+} (\xi ,\lambda )$ in different way. The probability to find the packet in the right well initially increases due to the tunnel transitions from the left well, and then returns to the initial value during the time $T_{rev} =\frac{2\pi }{(E_{1} -E_{0} )} \approx 300$. Meanwhile, a portion of the WP in the right well, which tunnels to the left well $U_{+}^{+} (\xi ,\lambda )$, is much smaller than in the reverse transitions, and the dynamics has typical oscillatory nature in the left well . This indicates that the partial trapping of the packet occurs in the deeper well, and typical oscillatory dynamics  is kept in the left well $U_{+}^{+} (\xi ,\lambda )$.



The initial distribution (\ref{susy-article-eqs:13}) could be sufficiently approximated by ten states of the  Hamiltonian $H_{+}^{+} $ (See table \ref{susy-article-tbl:2}), thus the proposed basis is effective enough for the description of localized states. Distributions like (\ref{susy-article-eqs:13}) are typical in the study of the  macroscopic tunneling of the Bose-Einstein condensate. Let's note that $\left|\Phi (\xi ,T)\right|$, calculated according to (\ref{susy-article-eqs:12}) coincides, with a good accuracy, with $\left|\Phi (\xi ,T)\right|=\left|\sum _{n=0}^{n_{\max } }c_{n} \, \psi _{\left(-\right)+}^{\left(-\right)+} (\xi ,E_{n} )\exp (-iE_{n} T) \right|$ where $c_{n}$ are chosen according to Table \ref{susy-article-tbl:2}.

\begin{table}[h]
\begin{footnotesize}
\begin{tabular}{|l||p{0.35in}|p{0.4in}|p{0.35in}|p{0.4in}|p{0.4in}|p{0.4in}|p{0.35in}|p{0.4in}|p{0.35in}|p{0.4in}|} \hline
\# of state & 0 & 1 & 2 & 3 & 4 & 5 & 6 & 7 & 8 & 9 \\ \hline \hline
$\lambda =0$ & 0.976 & 0 & 0.165 & 0 & 0 & 0.110 & 0 & 0.058 & 0 & -0.061 \\ \hline 
$\lambda =-0.95$ & 0.813 & -0.517 & 0.154 & -0.067 & 0.042 & -0.039 & 0.101 & 0.049 & 0.117 & 0.101 \\ \hline 
\end{tabular}
\end{footnotesize}
\caption{Coefficients of the expansion of initial wave packet (\ref{susy-article-eqs:13}).}
\label{susy-article-tbl:2}
\end{table}

\textbf{Large squeezed states}.

\begin{figure}
	    \begin{tabular}{cc}
	    a \epsfig{file=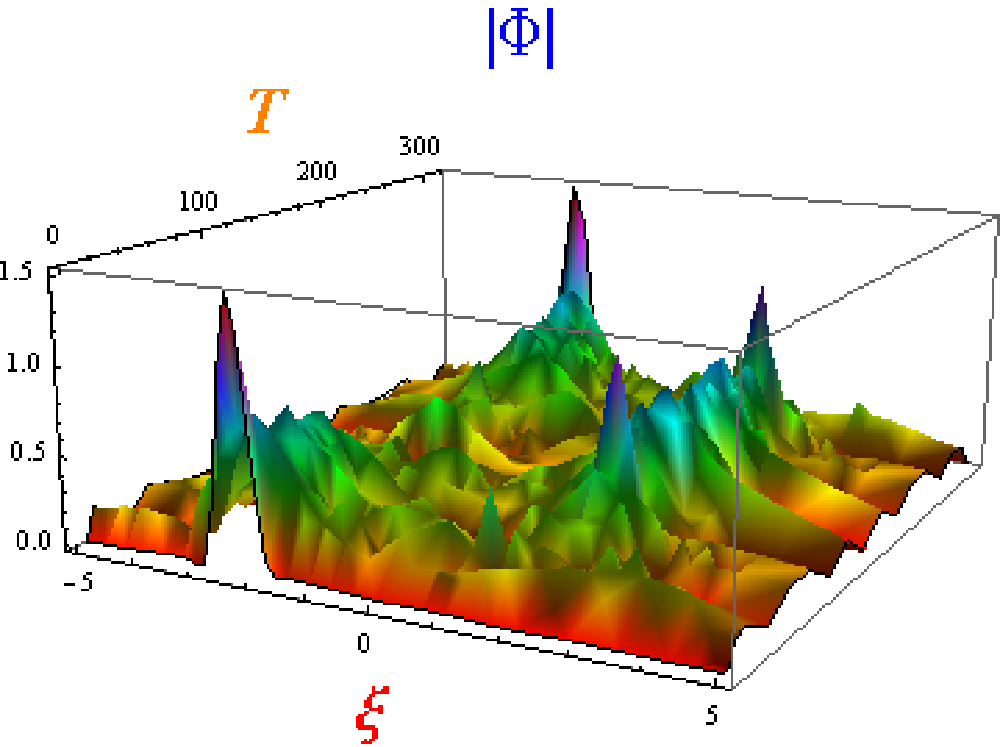, width=0.45\linewidth}
	    &
	    b \epsfig{file=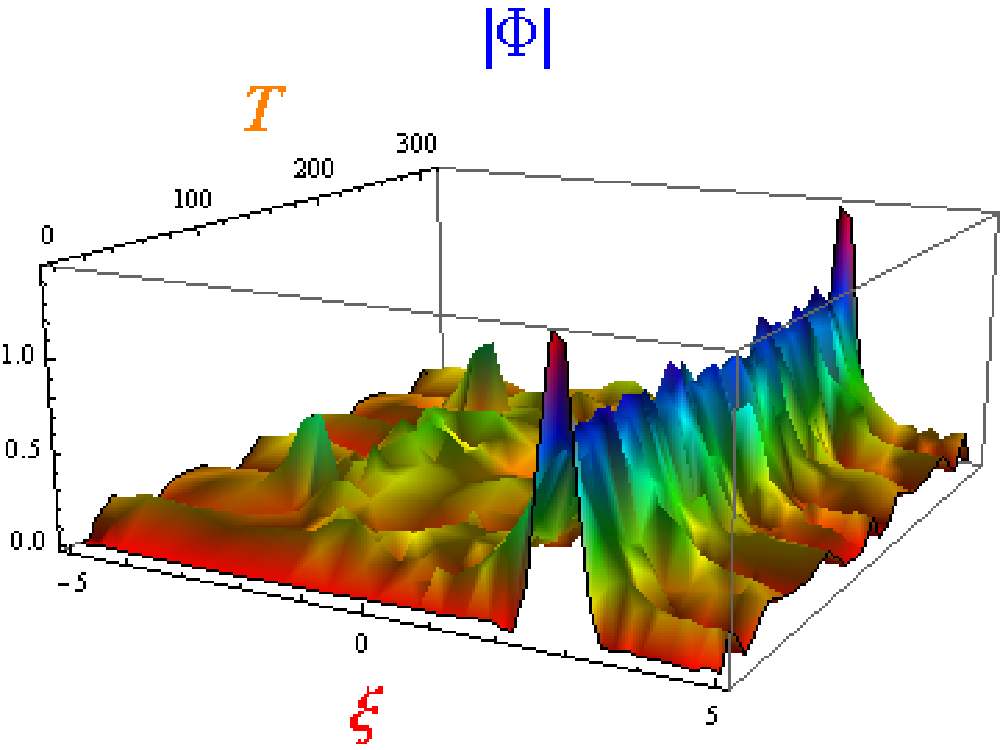, width=0.45\linewidth}
	    \end{tabular}
	    \caption{Time dependence of $\left|\Phi (\xi ,T)\right|$ for initial packet with $R=1.5$, localized in (a) -- left minimum of potential $U_{-}^{-} \left(\xi \right)$ $\left(\xi _{l} =-2.29\right)$, (b) -- in right minimum of $U_{+}^{+} \left(\xi ,\lambda \right)$, $\lambda =-0.95$ $(\xi _{l} =-2.153,~ \xi _{r} =2.755)$.}
	    \label{susy-article-fig:6}
     \end{figure}

When the degree of the wave packet localization increases, the number of states, which significantly contribute to $\Phi \left(\xi ,0\right)$, also increases. Let's consider the case when the initially localized state is located at one of the minima of the symmetric potential and in the deeper minimum of the asymmetric potential. When $R=1.5$ the WP with the center at $\xi _{0} =-2.29$ is sufficiently approximated by twenty states of the  Hamiltonian $H_{-}^{-} $ (for a symmetric potential). The similar localized state with the center located at $\xi _{0} =2.755$ needs twenty five states of $H_{+}^{+} $ for accurate approximation (for an asymmetric potential). Though the ground and the first excited states give the leading contribution to the expansion of wave packets, the contribution of higher states is still significant. It can be directly seen from the temporal dynamics of the initially localized states (Fig. \ref{susy-article-fig:6}).



The dynamics  in the symmetric potential is complex, and the process of the barrier crossing can not be even called tunneling. Such a dynamics could be observed in the case of the coherent tunneling breaking in periodically driven systems [23,24], e.g. the contribution of the over-barrier states completely mimics the tunneling of the under--barrier states. Nevertheless, the revival time of wave packets is still $T_{rev} =\frac{2\pi }{(E_{1} -E_{0} )} \approx 300$ according to the two-mode approximation. For the asymmetric potential the portion of the wave packet outside of the deeper minimum is small and its structure is  complex enough. At the same time, the wave packet in the global minimum squeezes at time scales $T\sim \frac{T_{rev} }{2} $ and restores at $T=T_{rev} $.



     \begin{figure}
	    \begin{tabular}{cc}
	    a \epsfig{file=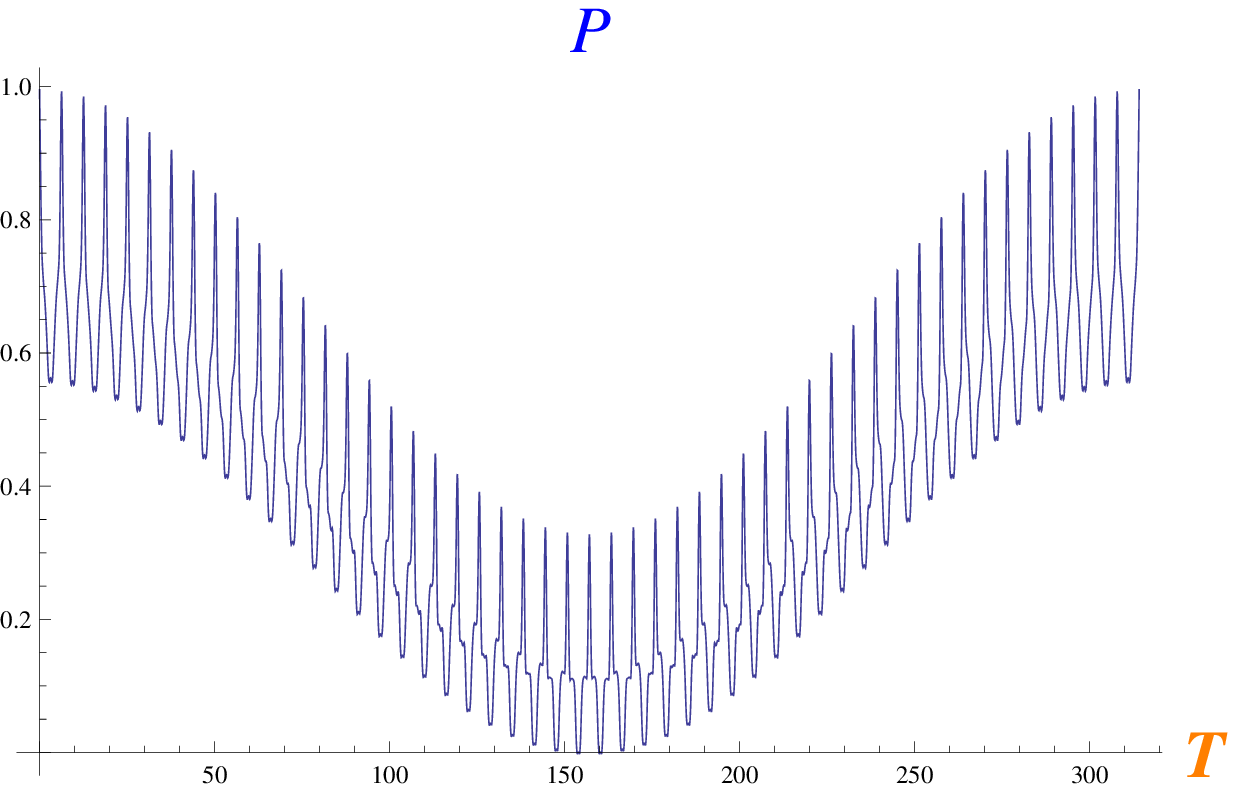, width=0.45\linewidth}
	    &
	    b \epsfig{file=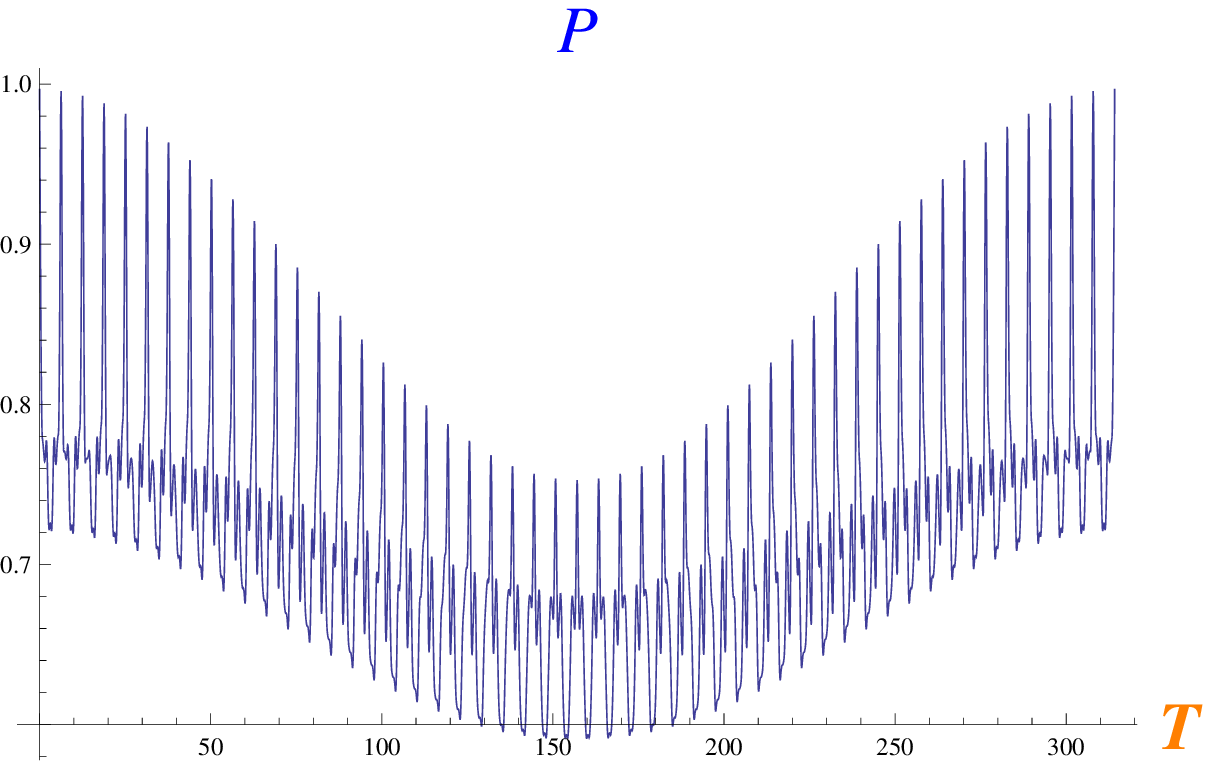, width=0.45\linewidth}
	    \end{tabular}
	    \caption{$P_{l(r)} (T)$ for the initial wave packet $\Phi \left(\xi ,0\right)$ with $R=1.5$ initially localized in (a) -- left minimum of the symmetric potential, (b) -- in right minimum of the asymmetric potential.}
	    \label{susy-article-fig:7}
     \end{figure}


This phenomena could be illustrated by the probability to find the particle in a certain well ям$P_{l(r)} (T)=\int _{x\in l(r)well}\left|\Phi (x,T)\right| ^{2} dx$ (Fig. \ref{susy-article-fig:7}). Large amplitudes of beats gives the evidence for the  substantial contribution from the over-barrier states to the dynamics of wave packets. As it was mentioned above, during the evolution in the symmetric potential, a substantial part of the wave packet leaves the left well and return to the initial value at the time scale $T_{rev} $. Meanwhile, the portion of the wave packet in the right well of the asymmetric potential varies insignificantly, that tells on the partial trapping of the wave packet.

\section{Conclusions}

 In the present paper we propose the approach to study of the dynamics of initially localized states, which is based on the exactly solvable quantum mechanical problems with multi-well potentials and on the corresponding exact propagators. Using the  Hamiltonian of the harmonic oscillator as a basis one, we obtain,  in frameworks of $N=4~SUSY~QM$, new Hamiltonians with multi--well potentials, both symmetric and asymmetric, together with the corresponding propagators. The study of the dynamics of the initially localized states demonstrates that the application of the two-mode approximation to the description of tunneling is very restricted, especially for systems with only few states in the under-barrier region. Such condition is typical for modern superconducting quantum interference devices and cold atoms traps. So, even the non--squeezed wave packet ($R=0$) can not be adequately approximated by wave functions of the ground and the first excited states.

 It is important to note that the states of Hamiltonians $H_{-}^{-} $ and $H_{+}^{+} $ are well--suited as the basic states to expand the localized states $\Phi \left(\xi ,0\right)$. Usually, ten states are enough for a good approximation of $\Phi \left(\xi ,0\right)$ in the sufficiently wide range of the squeezing parameter $R.$ This can be confirmed by a good agreement between $\left|\Phi (\xi ,T)\right|=\left|\sum _{n=0}^{n_{\max } }c_{n} \, \psi _{\left(-,+\right)}^{\left(-,+\right)} (\xi ,E_{n} )\exp (-iE_{n} T) \right|$ and the results of calculations by use of the exact propagator. In contrast to \cite{susy-article-ref:13, susy-article-ref:14, susy-article-ref:15, susy-article-ref:16}, where tunneling is compared in symmetric and asymmetric potentials, the spectra of Hamiltonians $H_{-}^{-} $ and $H_{+}^{+} $ are similar in our approach. Moreover, the shape of the potentials may be varied by the variation of $\bar{\varepsilon }$ and $\lambda $.

 The dynamics of WP contains slow tunneling of under--barrier states and fast beats, due to over-barrier states, and has a regular character, in contrast to \cite{susy-article-ref:13, susy-article-ref:15, susy-article-ref:16}. It happens since the over-barrier states have the spectrum of the initial Hamiltonian of HO (i.e. the equidistant spectrum) and the states get interfered with each other. Beats become smoother when a dissipation is taken into account \cite{susy-article-ref:25}. The revival time of the wave packet is equal to $T_{rev} =\frac{2\pi }{(E_{1} -E_{0} )} $, that coincides with revival time predicted by the two-mode approximation. When the squeezing parameter is low $\left(R=0.35\right)$ the amplitude of beats is relatively small, since the number of excited states, contributing to WP is low. In the symmetric potential a part of the WP, which is formed by tunneling of the under-barrier states is smooth enough.

 In the asymmetric potential the dynamics of WP, initially localized in one of the minima of $U_{+}^{+} \left(\xi ,\lambda \right)$, has a number of distinctive features. In particular, the partial trapping of WP in the initial well and the suppression of tunneling to another well is observed. This phenomenon occurs independently on the well initially containing the WP. When the initial state is uniformly distributed between both wells of $U_{+}^{+} \left(\xi ,\lambda \right)$, this phenomenon is observed for the deeper well, thus the tunneling rate from the deeper well to the other well is smaller than the rate of reverse transitions. When the squeezing parameter $R$ increases, beats, caused by the over-barrier states, increase and mimic the contribution of slow tunneling of the under-barrier states. In some sense it could be considered as the destruction of tunneling of the initially highly localized wave packet.

Authors thank to Yu.L. Bolotin for helpful discussions. We also thank to G.I. Ivashkevych for the help in preparation of the manuscript and to A.J. Nurmagambetov for careful reading and correcting the manuscript. This research was supported in part by the Joint DFFD--RFBR Grant \# F40.2/040.

\end{document}